\documentstyle[11pt,a4,ere99]{article}
\begin{document}

\title{The case for cosmic acceleration:\\ Inhomogeneity versus cosmological constant\\
\emph{Proceedings of the Spanish Relativity Meeting}\\ Bilbao, 1999}

\author{J.\-\ F.\ Pascual-S\'anchez}

\address{Dpto. Matem\'atica Aplicada Fundamental, \\Secci\'on Facultad de Ciencias, Universidad de Valladolid,\\
Valladolid, Spain\\E-mail: {\em {\tt jfpascua@maf.uva.es}}}  

\maketitle\abstracts{ In this work, I develop an alternative
explanation for the acceleration of the
 cosmic expansion, which seems to be a result of recent high redshift Supernova
  datae. In the current interpretation, this cosmic acceleration is explained by
  including a positive cosmological constant term (or vacuum energy), in the
   standard Friedmann models. Instead, I will consider a Locally Rotationally
    Symmetric (LRS) and spherically symmetric (SS), but inhomogeneous spacetime,
     with a barotropic perfect fluid equation of state for the cosmic matter.
     The congruence of matter has acceleration, shear and expansion.
     Within this framework the kinematical acceleration of the cosmic fluid
      or, equivalently, the inhomogeneity of matter, is just the responsible
      of the SNe Ia measured cosmic acceleration. Although in our model
       the Cosmological Principle is relaxed, it maintains almost isotropy
        about our worldline in agreement with CBR observations. }

\section{Introduction}
    Assuming the Cosmological Principle and hence the Friedmann (FLRW) models to be valid as global models of the Universe, for several decades astronomers have attempted to use the Hubble diagram of some predefined standard candle, to place constraints on the free parameters of the FLRW models, by comparing the observed redshift-luminosity distance relation at very low redshift (or alternatively the redshift-magnitude one) with that predicted in FLRW models with different values of the deceleration parameter $q_0$. Last year, two independent groups [1,2],
by using type Ia Supernovae as standard candles without evolution effects (but see recently [3]), were able to extend the Hubble diagram of luminosity distance versus redshift, out to a redshift of $z \stackrel{<}{{_\sim}} 1$, implementing a generalized K-correction.

    The main conclusion of these works is that the deceleration parameter at present cosmic time $q_0$, is negative, i.e., an acceleration of the cosmic expansion. As it is customary, they interpret this conclusion in the framework of the Friedmann (FLRW) models with cosmological constant, $\Lambda$, in which a necessary and sufficient condition of cosmic acceleration, if the weak energy condition holds, is that $\Lambda$ is positive.

    The cosmological constant $\Lambda$, was before reinterpreted as a vacuum energy and was used in the inflationary models. An estimate of this vacuum "quantum-mechanical" energy, is at least 120 orders of magnitude higher, that the vacuum energy associated with $\Lambda$, determined by the interpretation of the Supernova (SNe Ia) data in the background of FLRW models with $\Lambda$. It is not known which is the supression mechanism, if it exists.

    In this work, I develop an alternative explanation for the measured cosmic acceleration, which was first proposed in [4]. In this new explanation, from the beginning, $\Lambda$ and hence the vacuum energy are set to zero. Our starting point will be the relaxation of the essential assumption of the FLRW models, the Cosmological Principle and, after, the consideration of barotropic locally rotational symmetric (LRS) inhomogeneous models without $\Lambda$, in which the acceleration of the congruence of cosmic matter, which is related to the inhomogeneity of matter-energy, is just a sufficient condition for the SNe Ia measured cosmic acceleration.

    Our main "a priori" argument to consider inhomogeneous models [5], is observational.
Observationally, we can only assert that there is almost isotropy about our worldline and this has been falsified using different tests, being the most important one, the measured high degree of isotropy of the cosmic background radiation, CBR, in particular through use of the results of COBE and other posterior experiments. The exact isotropy about our worldline, when combined with the Copernican Principle, leads to exact isotropy about all worldlines (at late times, of different clusters of Galaxies and, at early times, of the average motion of a mixture of gas and radiation)
and thus to the exact homogeneity of the 3-dim spacelike hypersurfaces of constant cosmic time and finally to the FLRW models.

    However, as Ellis et al. [6] pointed out, if we suspend the Copernican assumption in favour of a direct observational approach, then it turns out that the measured almost isotropy of the CBR about our worldline, is insufficient to force exact isotropy into the spacetime geometry and hence exact spatial homogeneity of the 3-dim cosmic hypersurfaces, i.e., to force the verification of the Cosmological Principle.

Exact homogeneity of the 3-dim spacelike hypersurfaces have poor observational support. At the cosmological level, we only have data from our past light cone and testing homogeneity of the 3-dim hypersurfaces at constant global cosmic time, requires us to know about conditions at great distances at  present global cosmic time, whereas what we only can observe at great distances is what happened long time ago. Exact homogeneity  cannot be proven without either a fully determinate thoery of source evolution or availability of distance measures that are fully independent of the evolution of the suorces. So to test exact homogeneity of spacelike cosmic hypersurfaces,  we first have to understand how is the evolution of both the spacetime geometry and its matter-energy contents.

\section{Our model: Barotropic inhomogeneous spherically symmetric LRS}
The evidence for almost isotropy comes from the CBR and galaxy counts.
There is one family of spacetimes in which the Cosmological Principle is relaxed but they assure the observational almost isotropy, these are the locally rotationally symmetric (LRS)and spherically symmetric (SS) but spatially inhomogeneous models. In the family of inhomogeneous LRS spacetimes that we will consider, class IIc, the isometry group
is 3-dimensional, just half the isometry group of the FLRW models.

In our model, I will assume that the matter part of Einstein equations have a perfect fluid form. However, we will not consider the dust case, i.e., the Lemaitre-Tolman-Bondi (LTB) models, because then necessarily the congruence of matter worldlines will be geodesic or in free fall. Instead, I will consider a barotropic equation of state, $p=p(\varrho)$ and $\varrho+p>0$ (NEC condition),
which allows for an accelerating congruence.
If exact isotropy of CBR is assumed then the EGS theorem, assuming perfect fluid and expanding geodesic motion, uniquely select a FLRW spacetime. However, as Ferrando, Morales and Portilla showed [7], for a non-geodesic congruence or (and) an imperfect matter fluid, shear and vorticity free conformally stationary inhomogeneous spacetimes exist for which the CBR is exactly isotropic. In particular, this occurs in imperfect fluid LTB models or in the conformaly flat SS Stephani models (which do not admit a barotropic perfect fluid but they allow a thermodynamical scheme) which Dabrowski has recently considered to explain the SNe Ia datae [8]. But, note that almost isotropy allows the presence of shear as in the model considered.

LRS perfect fluid spacetimes have been before studied, for instance in [9] using the tetrad description and in [12], using as ours, the 1+3 threading formalism. Geometrically, in these LRS perfect fluid models (in particular class IIc in the Stewart-Ellis clasification,[9]), if one assumes spherical symmetry (SS), the coefficients of the spacetime metric depend on two independent variables of cosmic time and a radial coordinate, and if one chooses a comoving coordinate system,
then the metric depends on three non-negative coefficients, and reads
\begin{equation}\label{1}
ds^2= -N^2(r,t)\, dt^2 + B^2(r,t)\, dr^2 + R^2(r,t)\, d\Omega^2 .
\end{equation}

The congruence of matter fluid is initially irrotational and by the supposed barotropic equation of state (where by SS,  $\varrho=\varrho(r,t)$ and $p=p(r,t)$), the vorticity is zero at any time. However, the other kinematical quantities of the congruence of matter worldlines, i.e., acceleration, shear and expansion are non zero in this spacetime. Note that in the FLRW models all are zero except the expansion.

 As the vorticity of the matter flow and the spatial rotation (twist) of $e^1$ are zero, then the fluid matter flow is always hypersurface orthogonal and and $e^1$ is surface orthogonal, respectively, and there exists in our model: 1) A cosmic time function $t$, 2) A 3-metric of the spacelike hypersurfaces and 3) A spherical metric for the 2-dim surfaces. As far as I know, this spacetime was used by Mashhoon and Partovi to describe the gravitational collapse of a charged fluid sphere [10], and to obtain large-scale observational relations [11].

From the Einstein equations without $\Lambda$, one obtains the conservation of energy-momentum $T^{ab}$:
\begin{equation}\label{2}
\nabla_aT^{ab}=0   .
\end{equation}
From (\ref{2}), one obtains (see [15]) for a perfect fluid, the energy conservation equation:
\begin{equation}\label{3}
\frac{\partial \varrho}{\partial t}+ (\varrho +p)\,\Theta=0,
\end{equation}
being $\Theta$ the expansion of the matter fluid $\theta$ multiplied by the lapse $N$, and the Euler equation
\begin{equation}\label{4}
\frac{\partial p} {\partial r}+ (\varrho +p)\, a=0,
\end{equation}
being $a$, the acceleration of the fluid congruence. It should be
here emphasized that this kinematic acceleration is due to
pressure gradients, or equivalently, when a barotropic equation of
state is supposed, as in our work, to mass-energy gradients. This
kinematic acceleration is hence not origined by gravitation nor
inertia, which are, on the other hand, covariantly entangled in
General Relativity. This last error (the gravitational origin of
the acceleration) has been propagated through almost all the
literature on the subject.

Note that in FLRW models, equation (\ref{4}), is a tautology, because both terms on the LHS are independently zero. However,
the consequences of Euler equation (\ref{4}), are very important in our model. As the fluid is barotropic and the NEC holds, the acceleration is always away from a high-pressure region towards a neighbouring low-pressure one. In other words, the radial gradient of pressure is negative and gives place to an acceleration of the matter flow which opposes the gravitational attraction. This can also be important in order to surpass the classical singularity theorems, due to the fact that $\ddot S(t) > 0$ in our model, but in this work, I will only prove that this fluid acceleration can explain the SNe Ia data about the negativeness of the $q_0$ parameter.

\section{Luminosity distance-redshift relation and deceleration parameter}
To relate our model with the SNe Ia data, we need to know how the luminosity distance-redshift relation and the deceleration parameter are modified by the inhomogeneity. By using conservation of light flux, (see [14]), it follows from the metric (\ref{1})
\begin{equation}\label{5}
D_L=(1+z)^2R(t_s,r_s),
\end{equation}
being $D_L$ the luminosity distance and $t_s, r_s$ the cosmic time and radial coordinate at emission. At present time, $t_o$, this relation reads
\begin{equation}\label{6}
D_L(t_0,z)=(1+z)^2R[t_s(t_0,z),r_s(t_0,z)].
\end{equation}
If one makes an expansion of $D_L$ to second order in $z$, after making an expansion to first order in $z$ of $t_s(t_0,z)$ and $r_s(t_0,z)$, one finds, [4,11]:
\begin{equation}\label{7}
D_L(t_0,z) \approx \frac{1}{H_0}[z+\textstyle\frac{1}{2}(1-Q_0)z^2]  ,
\end{equation}
where $Q_0$ is a generalized deceleration parameter at present cosmic time.
On the other hand, if one develops the metric coefficients of (\ref{1}) and the mass-energy and pressure in power series of the radial coordinate and after imposing the Einstein equations, one obtains after a scale change in the radial coordinate, [4,11]:
\begin{eqnarray}\label{8}
ds^2 \approx& &- \left(1+\textstyle\frac{1}{2}\alpha(t) r^2\right)^2 dt^2                    \nonumber\\
          &  &+S^2(t)\left[\left(1+\textstyle\frac{1}{2}\beta(t) r^2\right)^2 dr^2 + r^2\left(1+\textstyle\frac{1}{2}\gamma(t) r^2\right)^2 d\Omega^2\right]  ,
\end{eqnarray}
where $S(t)$ is the usual scale factor, $\alpha(t)$ is a non-negative function related to the acceleration of the cosmic fluid and a combination of $\beta$ and $\gamma$ gives the intrinsic spatial curvature of the 3-dim spacelike cosmic spaces.

On the basis of equations (\ref{7},\ref{8}), one finds [4,11], that
\begin{equation}\label{10}
Q_0 = q_0 - I\!\!I_0 ,
\end{equation}
thus, the luminosity distance-redshift relation at present time, reads
\begin{equation}\label{11}
D_L(t_0,z) \approx \frac{1}{H_0}[z+\textstyle\frac{1}{2}(1-q_0+I\!\!I_0)z^2] ,
\end{equation}
where $H_0$ and $q_0$ are the usual Hubble and deceleration parameters
\[H_0:=\frac{\dot{S_0}}{S_0},\]
\[q_0:=-\frac{S_0 \ddot{S_0}}{\dot{S_0}^2},\]
and $I\!\!I_0$ is a new inhomogeneity parameter which reads
\begin{equation}\label{12}
I\!\!I_0= \frac{\alpha(t_0)}{(S_0H_0)^2}  .
\end{equation}
Note that  $I\!\!I_0$ is related to the congruence acceleration $A$, through the metric coefficient $\alpha(t)$.

In our model the deceleration parameter at present time is:
\begin{equation}\label{13}
q_0=\textstyle\frac{1}{2}\Omega_0\left( 1 + \displaystyle\frac{3p_0}{\varrho_0}\right)- I\!\!I_0,
\end{equation}
as at present time $\displaystyle\frac{3p_0}{\varrho_0}\ll 1$, then one finally obtains
\begin {equation}\label{14}
q_0 \approx\textstyle\frac{1}{2}\Omega_0- I\!\!I_0,
\end{equation}
where $\Omega_0$ is the present matter density in units of the critical density.

\section{Conclusions}
From the formulae (\ref{11}) and (\ref{14}), we see that one can obtain a negative deceleration parameter, i.e., cosmic acceleration, in agreement with recent SNe Ia data, by the presence of a positive inhomogeneity parameter related to the kinematic acceleration or, equivalently, to a negative pressure gradient or negative mass-energy gradient of the cosmic barotropic fluid. In this way, it is not necessary to explain the Supernova data by the presence of $\Lambda$ or a vacuum energy or some other exotic forms of matter. Although in our model without $\Lambda$, the Cosmological Principle is relaxed, however, it maintains perfect agreement with the almost isotropy about our worldline measured by the CBR observations.

\section*{Acknowledgments}
I am grateful to M.P. Dabrowski for drawing my attention to Stephani spacetimes and to A. San Miguel and F. Vicente for many discussions on this and (un)related subjects and TeX help. This work is partially supported by the spanish research projects VA61/98, VA34/99 of Junta de Castilla y Le\'on and C.I.C.Y.T. PB97-0487.

\vspace*{-2pt}
\section*{References}
[1] Perlmutter, S., et al., {\it Astrophys. J.}, {\bf 517} (1999) 565.\\
\noindent
[2] Riess, A.G., et al., {\it Astronomical J.}, {\bf 116} (1999) 1009.\\
\noindent
[3] Riess, A. G., preprint astro-ph/9907038, (1999).\\
\noindent
[4] Pascual-S\'anchez, J.-F., {\it Mod. Phys. Lett. A}, {\bf 14} (1999) 1539.\\
\noindent
[5] Krasi\'{n}ski, A., (1997) {\sl 'Inhomogeneous cosmological models'}, ed. C.U.P.\\
\noindent
[6] Ellis, G.F.R. et al., {\it Phys. Rep.}, {\bf 124} (1985) 315.\\
\noindent
[7] Ferrando, J. J., Morales, J.A., Portilla, M., {\it Phys. Rev. D}, {\bf 46} (1992) 578.\\
\noindent
[8] Dabrowski, M. P., preprint gr-qc/9905083, (1999).\\
\noindent
[9] Stewart, J.M., Ellis, G.F.R., {\it J. Math. Phys.}, {\bf9} (1968) 1072.\\
\hspace*{-.1in}
[10] Mashhoon, B., Partovi, M.H., {\it Phys. Rev. D}, {\bf 20} (1979) 2455.\\
\hspace*{-.1in}
[11] Partovi, M.H., Mashhoon, B., {\it Astrophys. J.}, {\bf 276} (1984) 4. \\
\hspace*{-.1in}
[12] Van Elst, H., Ellis, G.F.R., {\it Class. Quantum Grav.}, {\bf 13} (1996) 1099.\\
\hspace*{-.1in}
[13] Ellis, G.F.R., (1971) in {\sl 'General Relativity and Cosmology'},
                ed. R.K. Sachs (N.Y.: Academic Press).\\
\hspace*{-.1in}
[14] Kristian, J., Sachs, R.K. {\it Astrophys. J.}, {\bf 143} (1966) 379.\\
\hspace*{-.1in}

\end{document}